# Data-Driven Assessment of the County-Level Breast Cancer Incidence in the United States: Impacts of Modifiable and Non-Modifiable Factors

**Authors and affiliations:**
Tingting Zhao[1], Qing Han[2], Jinfeng Zhang[2]
[1] Department of Geography, Florida State University, 113 Collegiate Loop, Tallahassee FL 32306
[2] Department of Statistics, Florida State University, 117 N. Woodward Ave., Tallahassee FL 32306

**Corresponding authors:**
Tingting Zhao (tzhao@fsu.edu)
Jinfeng Zhang (jinfeng@stat.fsu.edu)

# Abstract

Female breast cancer (FBC) incidence rate (IR) varies greatly by counties across the United States (US). Factors responsible for such high spatial disparities are not well understood, making it challenging to design effective intervention strategies. We predicted FBC IRs using prevailing machine learning techniques for 1,754 US counties with a female population over 10,000. Outlier counties with the unexpectedly high or low FBC IRs were identified by controlling the non-modifiable factors (demographics and socioeconomics). Impacts of the modifiable factors (lifestyle, healthcare accessibility, and environment) were mapped. Our study also shed light on hidden FBC risk factors at the regional scale. Methods developed in our study may be used to discover the place-specific, population-level, modifiable factors for the intervention of other types of cancer or chronic diseases.

**Abbreviations**[1]

---

[1] Abbreviations:
DL - Deep Neural Networks
FBC - Female Breast Cancer
IR - Incidence Rate
KNN - K-Nearest Neighbor
MSE - Mean SquareEerror
RF - Random Forest
SVM - Support Vector Machine
XGBoost - Extreme Gradient Boost

# 1 Introduction

Female breast cancer (FBC) is the most common type of cancers in women in the United States (US). Approximately 129 new cases were diagnosed per 100,000 women across the country during the 2014-2018 time period (National Cancer Institute SEER Program, 2021). The FBC incidence rate (IR; cases per 100,000) varied greatly by counties according to the 2013-2017 SEER*Stat Database, ranging from 51.2 in Custer County, Colorado to 242.6 in Fallon County, Montana (National Cancer Institute, 2021).

What are the factors that might have caused the large variance of the county-level FBC IR across the US? At the patient level, FBC risk factors include age, genetic factors, reproductive history, taking hormones, and life style such as being physically inactive, being overweight, and drinking alcohol (Centers for Disease Controls and Prevention, 2021). At the population level, factors reported to have impacts on FBC incidence include, but not limited to, race (Li et al., 2020), access to health care (Murage et al., 2016, Broeders et al., 2018), socioeconomics (Buck, 2016, Zahnd and McLafferty, 2017), family composition (Zhao et al., 2020), environmental exposure (Hiatt and Brody, 2018), and neighborhood walkability and other characteristics (Gomez et al., 2015, Schootman et al., 2020). However, it is unclear how these individual- and population-level factors interact with one another and eventually influence the cancer outcomes manifested at the aggregated scales, such as counties or states (Mobley et al., 2012).

Among the documented risk factors, some are non-modifiable (such as a county's population racial composition), while others are modifiable (such as lifestyle or environment) (Maas et al., 2016, Cohen et al., 2023). For a county that saw a very high FBC IR, was it due mainly to its non-modifiable factors, or the performance of its modifiable factors, or some unknown factors at this aggregated scale? Knowing this would help the health professionals and policy makers to understand the challenges a county faces and to create intervention strategies targeting on modifiable aspects of counties with unexpectedly high breast cancer incidence.

In this study, we aimed to identify counties with significantly high or low FBC IRs compared to their expected IRs. The expected IRs are model estimates based on documented non-modifiable factors that cannot be modified easily through intervention strategies. Most of the demographic and socioeconomic factors belong to the non-modifiable category. This model allowed us to detect outlier counties, which were defined as those with actual IRs significantly higher or lower than the model predictions based on non-modifiable variables.

In addition, we created a second model to include both non-modifiable and modifiable factors. Most of the lifestyle, healthcare accessibility and environmental factors belong to the modifiable category. This second model allowed us to evaluate a county's performance of its modifiable factors in terms of curbing FBC incidence or increasing cancer detection rate. Lastly, we examined geographic patterns of the outlier counties, which help to identify potential unknown large-scale factors for FBC IRs. These results provide insights for local health department officials to investigate hidden contributing factors to the unexpectedly high or low FBC incidence in their counties.

# 2  Data and Methods

## 2.1 Data

The response variable in our study was the age-adjusted IR by counties for the 2011-2015 time period (United States Cancer Statistics, 2018), which is defined as the number of new cases per 100,000 females. An age-adjusted rate is a weighted average of the age-specific rates, where the weights are the proportions of persons in the corresponding age groups of the standard population. The predictor variables were drawn from demographic and socioeconomic data from the 2010 US Census (US Census Bureau, 2012) and the 2010-2015 5-Year American Community Survey (US Census Bureau, 2016). Also included were the lifestyle, healthcare, and environmental variables from the 2015 County Health Rankings National Data (Remington et al., 2015). In total, for 1,754 counties with a female population over 10,000, there were 93 explanatory variables in our dataset, which cover factors of demographics, socioeconomics, lifestyle, healthcare accessibility, and environment. All variables are known risk factors for FBC at the individual and/or population level (Hiatt et al., 2014, Zhao et al., 2020). These, however, excluded genetic (such as family history) or biological (such as breast density, age at menarche, and parity) factors that are not available at the county scale. All count data (such as Hispanic population) were normalized using percentage (such as percent Hispanic population among the entire population).

## 2.2 Model Fitting Using Machine Learning Algorithms

The machine learning algorithms we used in this study include Regularized regression method, LASSO; Random Forest (RF); Support Vector Machine (SVM); K-Nearest Neighbor (KNN); Extreme Gradient Boost (XGBoost); and Deep Neural Networks. Models such as LASSO, RF and XGBoost incorporate variable selection in their modeling procedure. The analyses were performed using R 4.0.3 with the caret, glmnet, randomForest, xgboost and e1071 packages as well as Python 3.7.1 with the keras and sklearn packages.

For model training, we used nested cross validation and grid search**.** Data were first randomly separated into 10 groups (outer folds). For each outer fold, we first set it aside and trained the model parameters on the other nine folds via an inner 10-fold cross validation process (the nine folds of data were separated into 10 inner folds), from which the best hyperparameters were found via grid search. With the best hyperparameters, the corresponding MSEs were calculated by comparing predicted IRs with actual IRs on each of the ten outer fold. The average of the 10 MSEs from all 10 folds was then calculated and used to measure the performance of a model.

For LASSO, variable selection was done using the shrinkage parameter, λ. For methods such as RF and XGBoost, variables were selected or ranked by importance score during the training process. Variables ranked in the top 30 in terms of importance scores (for RF and XGBoost) or selected by the shrinkage parameter (for LASSO) were kept for each training process. Through this step, we were able to screen out unimportant factors to reduce noise and prevent over-fitting. As a result, the variables chosen by all ten inner folds' training-validation processes were identified as the most influential variables by that model. The model-selected variables were then combined with additional variables selected through expert knowledge to include well-known cancer risk factors (Appendix A). This combined set was subject to the second round of hyperparameter tuning, again using the ten inner folds. The hyperparameters after this tuning would constitute the ultimate prediction models. This variable selection and parameter tuning process was done for each of the 10 outer folds.

### 2.3 Identifying Outlier Counties

The outlier counties were identified based on predictions of the best performance model, LASSO, with only non-modifiable factors as the model inputs. The model residual was calculated as the observed value subtracting the predicted value (Equation 1). A negative residual indicates that the observed actual IR is lower than expectation.

$$e_i = y_i - \hat{y}_i \qquad (1)$$

where $y_i$ stands for the observed IR for a particular county $i$, and $\hat{y}_i$ stands for the predicted value by the model for county $i$.

Anselin Local Moran's I (Anselin, 1995) was applied to the residuals calculated for the LASSO model. This spatial analysis uses z-score to identify spatial clustering of the high (or low) residual values. A high positive z-score indicates that a county and its surrounding counties have similar values of model residuals. It helps identify statistically significant cluster of outlier counties. This analysis was performed with the Cluster and Outlier Analysis in ArcGIS Pro 2.4.0. For the parameters of spatial relationships, we examined contiguity edges corners (for which only the adjacent counties are counted as neighboring counties); fixed distance band (with multiple distance thresholds), and inverse distance (that include all counties within certain distance thresholds as neighboring counties).

### 2.4 Identifying Counties' Performance on Modifiable Factors

This analysis involved comparing predictions of model 1 and model 2 (Equation 2). Model 1 is the LASSO prediction based on non-modifiable factors only. Model 2 is the LASSO prediction based on both non-modifiable and modifiable factors. The difference of the model 1 and 2's predicted IRs was used as the indicator for a county's performance of modifiable factors as a whole. A relatively large negative or positive value for a county indicates that the modifiable factors significantly affected its FBC IR.

$$d_i = \hat{y}_2 - \hat{y}_1, \qquad (2)$$
where $\hat{y}_1$ is the predicted IR by model 1 and $\hat{y}_2$ is the predicted IR by model 2.

# 3 Results

## 3.1 Model Performance

Among all the models, LASSO performed the best in terms of MSE and correlation coefficient (Table 1). The mean square error (MSE) measures differences between the observed Age-adjusted IRs and the predicted ones. The lower the value is, the closer the model predictions are to the true values. Correlation coefficient is another metrics that measures the linear relationships between the predicted values and the true values. A correlation coefficient value closer to 1 represents a stronger association between the predicted and true values. R-squared values were also calculated and they generally agree with those of MSEs and correlation coefficients.

Table 1. Performance of models

|  | Metrics | BIC | LASSO | RF | SVM | KNN | XGBoost | DL |
|---|---|---|---|---|---|---|---|---|
| Model 1 | MSE | 172.685 (18.821) | 167.205 (16.451) | 173.870 (17.185) | 171.088 (14.553) | 178.657 (17.846) | 173.412 (16.803) | 177.241 (19.655) |
|  | Correlation | 0.570 (0.042) | 0.587 (0.035) | 0.563 (0.038) | 0.576 (0.035) | 0.554 (0.047) | 0.569 (0.043) | 0.548 (0.051) |
|  | $R^2$ | 0.327 (0.048) | 0.345 (0.042) | 0.318 (0.042) | 0.333 (0.041) | 0.309 (0.052) | 0.325 (0.048) | 0.297 (0.058) |
| Model 2 | MSE | 165.229 (14.718) | 162.285 (15.075) | 162.663 (14.748) | 163.133 (13.633) | 167.876 (19.957) | 167.199 (15.836) | 183.000 (16.191) |
|  | Correlation | 0.594 (0.032) | 0.602 (0.034) | 0.602 (0.043) | 0.602 (0.038) | 0.591 (0.054) | 0.590 (0.044) | 0.559 (0.036) |
|  | $R^2$ | 0.354 (0.038) | 0.364 (0.042) | 0.364 (0.052) | 0.364 (0.046) | 0.351 (0.061) | 0.350 (0.052) | 0.270 (0.071) |

## 3.2 Final Model for Prediction

The best model that can be used for future county-level FBC IR prediction was from LASSO with coefficients (Appendix A). Other machine learning models achieved similar performance using variables largely overlap with those of LASSO.

The variables selected by LASSO model provide some insight on the most influential factors for the county-level FBC IR prediction (Appendix A). Race, urban/rural setting, poverty rates, income, and education attainment are the most important factors among those in the socioeconomic category. Family composition, especially the presence of older children, is a significant factor for the county-level FBC IR. Percent of the elderly receiving mammography and rates of the insured population are the top contributors in the access to health care category. Lifestyle factors such as smoking and commuting time were also found as predictors.

## 3.3 Outlier Counties

Using the Anselin Local Moran's I (based on the inverse distance weighting of counties within 100 kilometers), we identified a few spatial clusters of counties with unexpectedly low IRs in south Michigan, west of Virginia, south Florida, and Texas (Figure 1). These counties' observed IRs were significantly lower than their expected IRs based on non-modifiable factors. Clusters of unexpectedly high-IR counties were found in west New York, New Hampshire, North Carolina, central Kentucky, and central Alabama (Figure 1). These counties' observed IRs were significantly higher than the expected IRs based on non-modifiable factors.

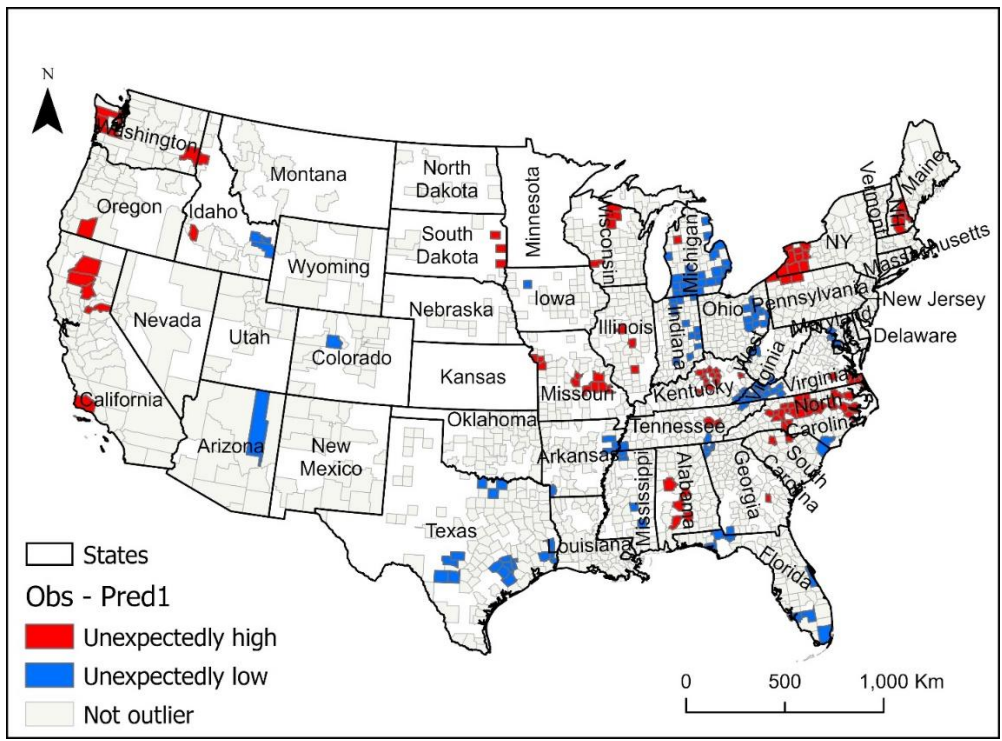

Figure 1. Outlier counties across the US. *Obs*: observed FBC IR. *Pred1*: predicted FBC IR based on non-modifiable factors. *Unexpectedly high*: counties where the observed IR is significantly higher than the model prediction. *Unexpectedly how*: counties where the observed IR is significantly lower than the model prediction.

We also mapped the impacts of modifiable factors (such as lifestyle and health accessibility) on counties' FBC IRs (Figure 2). Clusters of counties in west Pennsylvania-Ohio, south Florida, north Mississippi, west Louisiana, and central Oklahoma saw a reduced IRs due to their modifiable factors. In contrast, counties in west Michigan, north Illinois-south Wisconsin, North Carolina, west Maryland-Delaware, and Massachusetts were found to have increased IRs due to their modifiable factors.

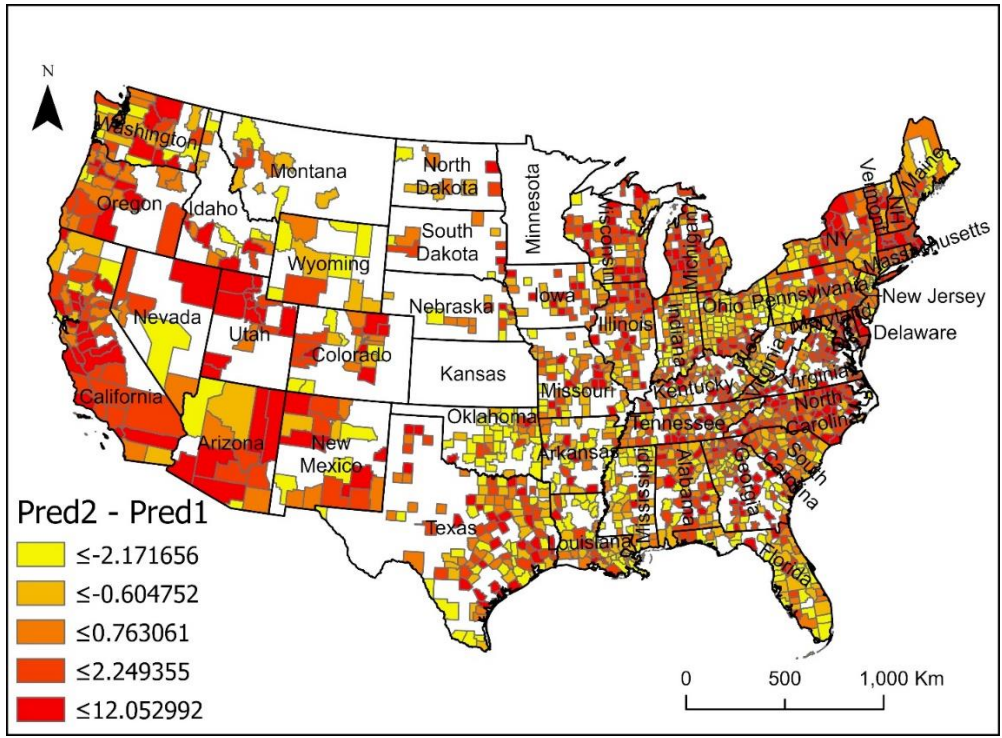

Figure 2. Impacts of the modifiable factors on counties' IRs. Negative value indicates a county's IR was reduced by its modifiable factors. Positive value indicates a county's IR was increased by its modifiable factors. *Pred1*: predicted FBC IR based on non-modifiable factors. *Pred2*: predicted FBC IR based on both non-modifiable and modifiable factors.

## 4 Discussion

### 4.1 Prediction of FBC Incidence

The impacts of FBC on the population and challenges associated with FBC prevention have fueled studies of cancer outcomes prediction, which in turn provide insights for governments, policy makers, and health professionals. While models have been developed extensively to assess risk factors for individuals' probability to develop FBC (Meads et al., 2012, Henley et al., 2020), risk prediction is still challenging at the population level (Howell et al., 2014). Many of the models based on population-level data were driven solely by the population growth and aging (Bray and Moller, 2006). Some recent models also included race as well as factors such as reduction in the use of hormone replacement therapy (Weir et al., 2015) and urban-rural settings (Katayama and Narimatsu, 2016).

Models incorporating multiple factors to predict the county-level FBC IRs are still underdeveloped, though many of the patient-level risk factors were reported as significant indicators of cancer outcomes at the population level. In a recent study, the ratio of mortality over incidence was examined for all cancers combined across counties in the US; and the author identified significant contributive variables such as income, education, medical assess, race, marriage, number of dependents, obesity, and poor health etc. (Buck, 2016). Obesity was found significantly positively associated with breast cancer mortality, while mammogram screening was negatively associated with the latter (Anderson et al., 2023). Another research group analyzed incidence of inflammatory breast cancer, a rare form of breast cancer. They found that race, poverty, and urbanicity differed significantly for the high vs. low cluster counties across the US (Scott et al., 2017). Most recently, the county-level FBC incidence by racial/ethnic groups was linked to a county's percentage of married-couple families, median income, accessibility to health care, and air pollution measured as $PM_{2.5}$ (Zhao et al., 2020).

As machine learning methods become mainstream for predicting disease outcomes, more and more studies started utilizing this technique to predict cancer outcomes and identify significant contributing factors. Researchers compared eight machine learning algorithms using seven metrics on cancer detection (Goldberger, 1991). The dataset was rather small with only 116 subjects and ten quantitative features. Another study implemented 11 algorithms on 116 patients and nine features (Moore et al., 2016). In 2016, a patient-level study compared four algorithms on 699 individuals (Benign: 458 Malignant: 241) and 11 integer-valued attributes, and reported that Support Vector Machine (SVM) gave the highest accuracy (Jung et al., 2007).

These previous studies have relatively small datasets, which may not provide an unbiased comparison of different machine learning methods since many machine learning methods require a decent amount of training data. Smaller sample sizes also lead to less accurate models and less reliable prediction results. In this study, we used 1,754 counties' FBC IR data and 93 variables in our analysis, which is the largest scale of such study as far as we know. The nested 10-fold cross-validation for training and evaluation and grid search for hyperparameters tuning used in this study are well-established protocols in machine learning field to ensure we obtain best performing models. LASSO performed the best among all the machine learning methods (by narrow margins). This is somewhat surprising since in the modern machine learning field, more sophisticated methods, such as RF and XGboost usually perform better, as shown by many of the public challenges hosted on Kaggle. We hypothesize that the data may be inherently linear and the noise level in the data may also play a role since regularized linear regression models are the most robust ones among all the models used in the study.

It is worth noting that, as the observed IR for a county is the product of its actual IR and the detection rate, a high value of IR is not necessarily bad if it is caused by a higher rate of cancer detection. The goals of health agencies are to reduce the actual IR while increasing the detection rate. In this study, we focused on developing a framework that uses advanced machine learning methods to compare IRs of different counties by separating modifiable and non-modifiable factors. This allows researchers and health policy makers to identify differences caused by modifiable factors, which in turn will help design better strategies to either reduce the actual IR or increase the detection rate. It would be interesting to further separate modifiable factors into cancer detection factors and cancer risk factors. This is the subject of our future studies.

## 4.2 County-Level FBC Risk Factors

Our study indicates that, among the non-modifiable factors, counties with a higher percentage of White population saw higher FBC IR, which corroborates with previous research findings of racial disparity of FBC IR (Seiler et al., 2017, Li et al., 2020). In addition, urban areas, higher female education, lower poverty rate, and higher income are associated with higher FBC IR. FBC incidences have been higher in affluent classes with high education attainment and low poverty rates (Klassen and Smith, 2011, Scott et al., 2017). Among the modifiable factors, increasing mammography correlates with the higher chance of diagnosing FBC (Gangnon et al., 2015). Likewise, the percent of insured people and PCP rates are also indicators for accessibility to health care (Kirby et al., 2017) that facilitates the detection of cancer. The percent of adults who smoke was also identified as a positive factor related to the county-level FBC IR.

Additionally, we found two new risk factors for a county's FBC IR. First, the presence of children at home was found as a significant factor, adding complication to the impacts of family characteristics on FBC IR reported in an earlier study (Zhao et al., 2020). In particular, it is not very clear why a higher proportion of families with older children correlates with a higher county-level FBC IR. The percent of biological children that makes up family households was found to decrease the FBC IR. This perhaps relates to parity (i.e., number of pregnancies carried by a woman for at least 20 weeks) that has been documented to decrease FBC risk with varying impacts across age groups (Albrektsen et al., 1994, Lima et al., 2020). The number of biological children may also be negatively correlated with the use of oral contraceptives, which was reported to increase FBC risk (Zhu et al., 2012).

The second new county-level factor was the percent of long-driving commuters contributes positively to the county-level FBC IR (Appendix A). This has not been very well documented in the past at the patient or population level. The percentage of long-driving commuters is a measure of the workforce that drives alone for 30 minutes or longer to work (University of Wisconsin Population Health Institute, 2015). A recent systematic review and meta-analysis of active commuting's impacts on cancer incidence indicated no association between the two (Dinu et al., 2019). However, cycling compared to car or public transportation was reported to reduce the cancer incidence significantly in UK (Celis-Morales et al., 2017). A more recent longitudinal study of the England and Wales cohort reported an 11% reduced cancer IR in bicycle commuters compared to those commuting by private motorized vehicle (Patterson et al., 2020).

## 4.3 Counties with Unexpected FBC IRs

Disease mapping is not uncommon for breast cancer thanks to the rapid growth of spatial epidemiology (Roquette et al., 2017, Schootman et al., 2017). Our spatial outlier analyses is different from the traditional mapping of geographic clusters of high or low cancer incidence (Mandal et al., 2009, Tatalovich et al., 2015, Scott et al., 2017), where only patterns of cases or IR were examined without control for the influencing factors such as those in our models. The spatial patterns of cancer incidence were significantly affected by the covariates used in the analysis (Vieira et al., 2002, Goovaerts, 2010, Kuo et al., 2011, Wang et al., 2012).

Our maps reveal specific information for policy implications and guide future research in breast cancer prevention, which was not provided by previous mapping studies. In our study, the modifiable and non-modifiable factors were separated for the purpose of evaluating a county's performance on FBC prevention. Counties' performance of the modifiable factors (Figure 2) should be the ones that health professionals and policy makers examine closely. These may be the counties excel in cancer detection through healthcare accessibility or insurance availability (Jayasekera et al., 2019) or to be improved through the intervention of lifestyle and/or environmental factors to curb their FBC incidence. As discussed earlier, whether a county actually did well or poorly in FBC prevention will needed to be investigated by examining the individual modifiable factors instead of simply judging from the difference between the observed and predicted IRs.

In addition, the spatial patterns of the two model outputs help scholars to form new research questions about unknown contextual factors for FBC. For example, one of the high-incidence clusters was in the western New York (Figure 1). However, the known modifiable factors were found to decrease those counties' FBC IRs (Figure 2). What other unknown factors may have contributed to the excessively high IRs for this cluster? According to literature, the home county of Buffalo and a few nearby counties were identified as one of the most likely high FBC death clusters during the 1988-1992 time period (Kulldorff et al., 1997). In addition, a population-based case-control study documented a spatially clustered pattern of residences at birth for pre-menopausal cases (1996-2001) in Erie and Niagara counties. It was suspected that the environmental exposure to compounds such as PAHs and benzene in those residential areas may have caused an increased FBC incidence (Han et al., 2004). Similar findings were reported in counties of southern Ontario, Canada, where association with industrial pollutants was suggested as a potential cause (Luginaah et al., 2012). Based on these literature findings, it is safe to speculate that pollution is likely the underlying factor contributing to the cluster of the unexpectedly high incidence of breast cancer in these western-New York counties.

## 5. Conclusions

Breast cancer incidence varies greatly by counties across the US. Factors contributing to the large spatial variance are not well understood, making it challenging to design effective intervention strategies targeting modifiable factors. In this study, we separated non-modifiable and modifiable factors to predict the expected FBC IRs for 1,754 US counties with a female population over 10,000. The non-modifiable factors included demographics and socioeconomics, while modifiable factors included lifestyle, access to healthcare, and environmental variables. Prevailing machine learning techniques, including LASSO, Support Vector Machine, K-Nearest Neighbors, Random Forest, XGBoost, and deep Neural Networks, were applied and compared.

We found that LASSO performed the best in predicting the county-level FBC IRs. Significant predictors included race, urban/rural setting, poverty rates, income, education, family composition, rates of mammography, insured population, smoking, and commuting time. Our method provides a framework to identify counties with the unexpectedly high or low FBC IRs by controlling their non-modifiable factors. The impacts of their modifiable factors were also mapped to investigate whether they are the potential causal factors for the unexpectedly high or low FBC IRs. Additionally, our study shed light on hidden FBC risk factors at the regional scale.

The fact that not all of the important factors are modifiable makes it complicated to identify counties where IRs can be reduced through changing the modifiable factors. Our study developed a framework to separate impacts of the modifiable factors to provide better guidance on counties' future breast cancer intervention strategies. This framework can be applied to other spatial scales, for example, states or health service areas. It may also be used to study other types of cancer or chronic diseases.

# References

ALBREKTSEN, G., HEUCH, I., TRETLI, S. & KVÅLE, G. 1994. Breast Cancer Incidence before Age 55 in Relation to Parity and Age at First and Last Births: A Prospective Study of One Million Norwegian Women. *Epidemiology,* 5.

ANDERSON, T., HERRERA, D., MIREKU, F., BARNER, K., KOKKINAKIS, A., DAO, H., WEBBER, A., MERIDA, A. D., GALLO, T. & PIEROBON, M. 2023. Geographical Variation in Social Determinants of Female Breast Cancer Mortality Across US Counties. *JAMA network open,* 6**,** e2333618-e2333618.

ANSELIN, L. 1995. LOCAL INDICATORS OF SPATIAL ASSOCIATION - LISA. *Geographical Analysis,* 27**,** 93-115.

BRAY, F. & MOLLER, B. 2006. Predicting the future burden of cancer. *Nature Reviews Cancer,* 6**,** 63-74.

BROEDERS, M. J. M., ALLGOOD, P., DUFFY, S. W., HOFVIND, S., NAGTEGAAL, I. D., PACI, E., MOSS, S. M. & BUCCHI, L. 2018. The impact of mammography screening programmes on incidence of advanced breast cancer in Europe: a literature review. *BMC Cancer,* 18**,** 860.

BUCK, K. D. 2016. Modelling of geographic cancer risk factor disparities in US counties. *Applied Geography,* 75**,** 28-35.

CELIS-MORALES, C. A., LYALL, D. M., WELSH, P., ANDERSON, J., STEELL, L., GUO, Y., MALDONADO, R., MACKAY, D. F., PELL, J. P., SATTAR, N. & GILL, J. M. R. 2017. Association between active commuting and incident cardiovascular disease, cancer, and mortality: prospective cohort study. *BMJ,* 357**,** j1456.

CENTERS FOR DISEASE CONTROLS AND PREVENTION. 2021. *What Are the Risk Factors for Breast Cancer?* [Online]. Available: https://www.cdc.gov/cancer/breast/basic_info/risk_factors.htm [Accessed June 18 2021].

COHEN, S. Y., STOLL, C. R., ANANDARAJAH, A., DOERING, M. & COLDITZ, G. A. 2023. Modifiable risk factors in women at high risk of breast cancer: a systematic review. *Breast Cancer Research,* 25**,** 45.

DINU, M., PAGLIAI, G., MACCHI, C. & SOFI, F. 2019. Active Commuting and Multiple Health Outcomes: A Systematic Review and Meta-Analysis. *Sports Medicine,* 49**,** 437-452.

GANGNON, R. E., SPRAGUE, B. L., STOUT, N. K., ALAGOZ, O., WEEDON-FEKJAER, H., HOLFORD, T. R. & TRENTHAM-DIETZ, A. 2015. The Contribution of Mammography Screening to Breast Cancer Incidence Trends in the United States: An Updated Age-Period-Cohort Model. *Cancer Epidemiology Biomarkers & Prevention,* 24**,** 905-912.

GOLDBERGER, A. S. 1991. *A course in econometrics*, Harvard University Press.

GOMEZ, S. L., SHARIFF-MARCO, S., DEROUEN, M., KEEGAN, T. H. M., YEN, I. H., MUJAHID, M., SATARIANO, W. A. & GLASER, S. L. 2015. The impact of neighborhood social and built environment factors across the cancer continuum: Current research, methodological considerations, and future directions. *Cancer,* 121**,** 2314-2330.

GOOVAERTS, P. 2010. Visualizing and testing the impact of place on late-stage breast cancer incidence: A non-parametric geostatistical approach. *Health & Place,* 16**,** 321-330.

HAN, D., ROGERSON, P. A., NIE, J., BONNER, M. R., VENA, J. E., VITO, D., MUTI, P., TREVISAN, M., EDGE, S. B. & FREUDENHEIM, J. L. 2004. Geographic clustering of residence in early life and subsequent risk of breast cancer (United States). *Cancer Causes & Control,* 15**,** 921-929.

HENLEY, S. J., WARD, E. M., SCOTT, S., MA, J., ANDERSON, R. N., FIRTH, A. U., THOMAS, C. C., ISLAMI, F., WEIR, H. K., LEWIS, D. R., SHERMAN, R. L., WU, M., BENARD, V. B., RICHARDSON, L. C., JEMAL, A., CRONIN, K. & KOHLER, B. A. 2020. Annual report to the nation on the status of cancer, part I: National cancer statistics. *Cancer,* 126**,** 2225-2249.

HIATT, R. A. & BRODY, J. G. 2018. Environmental Determinants of Breast Cancer. *Annu Rev Public Health,* 39**,** 113-133.

HIATT, R. A., PORCO, T. C., LIU, F. C., BALKE, K., BALMAIN, A., BARLOW, J., BRAITHWAITE, D., DIEZ-ROUX, A. V., KUSHI, L. H., MOASSER, M. M., WERB, Z., WINDHAM, G. C. & REHKOPF, D. H. 2014. A Multilevel Model of Postmenopausal Breast Cancer Incidence. *Cancer Epidemiology Biomarkers & Prevention,* 23**,** 2078-2092.

HOWELL, A., ANDERSON, A. S., CLARKE, R. B., DUFFY, S. W., EVANS, D. G., GARCIA-CLOSAS, M., GESCHER, A. J., KEY, T. J., SAXTON, J. M. & HARVIE, M. N. 2014. Risk determination and prevention of breast cancer. *Breast Cancer Research,* 16.

JAYASEKERA, J., ONUKWUGHA, E., CADHAM, C., HARRINGTON, D., TOM, S., PRADEL, F. & NASLUND, M. 2019. An ecological approach to monitor geographic disparities in cancer outcomes. *Plos One,* 14.

JUNG, I., KULLDORFF, M. & KLASSEN, A. C. 2007. A spatial scan statistic for ordinal data. *Stat Med,* 26**,** 1594-607.

KATAYAMA, K. & NARIMATSU, H. 2016. Prediction of Female Breast Cancer Incidence among the Aging Society in Kanagawa, Japan. *Plos One,* 11.

KIRBY, R. S., DELMELLE, E. & EBERTH, J. M. 2017. Advances in spatial epidemiology and geographic information systems. *Annals of Epidemiology,* 27**,** 1-9.

KLASSEN, A. C. & SMITH, K. C. 2011. The enduring and evolving relationship between social class and breast cancer burden: A review of the literature. *Cancer Epidemiology,* 35**,** 217-234.

KULLDORFF, M., FEUER, E. J., MILLER, B. A. & FREEDMA, L. S. 1997. Breast Cancer Clusters in the Northeast United States: A Geographic Analysis. *American Journal of Epidemiology,* 146**,** 161-170.

KUO, T. M., MOBLEY, L. R. & ANSELIN, L. 2011. Geographic disparities in late-stage breast cancer diagnosis in California. *Health & Place,* 17**,** 327-334.

LI, Y., PANG, X., CUI, Z., ZHOU, Y., MAO, F., LIN, Y., ZHANG, X., SHEN, S., ZHU, P., ZHAO, T., SUN, Q. & ZHANG, J. 2020. Genetic factors associated with cancer racial disparity – an integrative study across twenty-one cancer types. *Molecular Oncology,* 14**,** 2775-2786.

LIMA, S. M., KEHM, R. D., SWETT, K., GONSALVES, L. & TERRY, M. B. 2020. Trends in Parity and Breast Cancer Incidence in US Women Younger Than 40 Years From 1935 to 2015. *JAMA network open,* 3**,** e200929-e200929.

LUGINAAH, I. N., GOREY, K. M., OIAMO, T. H., TANG, K. X., HOLOWATY, E. J., HAMM, C. & WRIGHT, F. C. 2012. A geographical analysis of breast cancer clustering in southern Ontario: generating hypotheses on environmental influences. *International Journal of Environmental Health Research,* 22**,** 232-248.

MAAS, P., BARRDAHL, M., JOSHI, A. D., AUER, P. L., GAUDET, M. M., MILNE, R. L., SCHUMACHER, F. R., ANDERSON, W. F., CHECK, D., CHATTOPADHYAY, S., BAGLIETTO, L., BERG, C. D., CHANOCK, S. J., COX, D. G., FIGUEROA, J. D., GAIL, M. H., GRAUBARD, B. I., HAIMAN, C. A., HANKINSON, S. E., HOOVER, R. N., ISAACS, C., KOLONEL, L. N., LE MARCHAND, L., LEE, I. M., LINDSTROM, S., OVERVAD, K., ROMIEU, I., SANCHEZ, M. J., SOUTHEY, M. C., STRAM, D. O., TUMINO, R., VANDERWEELE, T. J., WILLETT, W. C., ZHANG, S. M., BURING, J. E., CANZIAN, F., GAPSTUR, S. M., HENDERSON, B. E., HUNTER, D. J., GILES, G. G., PRENTICE, R. L., ZIEGLER, R. G., KRAFT, P., GARCIA-CLOSAS, M. & CHATTERJEE, N. 2016. Breast Cancer Risk From Modifiable and Nonmodifiable Risk Factors Among White Women in the United States. *Jama Oncology,* 2**,** 1295-1302.

MANDAL, R., ST-HILAIRE, S., KIE, J. G. & DERRYBERRY, D. 2009. Spatial trends of breast and prostate cancers in the United States between 2000 and 2005. *International Journal of Health Geographics,* 8**,** 53.

MEADS, C., AHMED, I. & RILEY, R. D. 2012. A systematic review of breast cancer incidence risk prediction models with meta-analysis of their performance. *Breast Cancer Research and Treatment,* 132**,** 365-377.

MOBLEY, L. R., KUO, T. M., WATSON, L. & BROWN, G. G. 2012. Geographic disparities in late-stage cancer diagnosis: Multilevel factors and spatial interactions. *Health & Place,* 18**,** 978-990.

MOORE, S. C., LEE, I. M., WEIDERPASS, E., CAMPBELL, P. T., SAMPSON, J. N., KITAHARA, C. M., KEADLE, S. K., AREM, H., DE GONZALEZ, A. B., HARTGE, P., ADAMI, H. O., BLAIR, C. K., BORCH, K. B., BOYD, E., CHECK, D. P., FOURNIER, A., FREEDMAN, N. D., GUNTER, M., JOHANNSON, M., KHAW, K. T., LINET, M. S., ORSINI, N., PARK, Y., RIBOLI, E., ROBIEN, K., SCHAIRER, C., SESSO, H., SPRIGGS, M., VAN DUSEN, R., WOLK, A., MATTHEWS, C. E. & PATEL, A. V. 2016. Association of Leisure-Time Physical Activity With Risk of 26 Types of Cancer in 1.44 Million Adults. *Jama Internal Medicine,* 176**,** 816-825.

MURAGE, P., CRAWFORD, S. M., BACHMANN, M. & JONES, A. 2016. Geographical disparities in access to cancer management and treatment services in England. *Health & Place,* 42**,** 11-18.

NATIONAL CANCER INSTITUTE. 2021. *State Cancer Profiles: Incidence Rates Table* [Online]. Available: https://www.statecancerprofiles.cancer.gov/incidencerates/ [Accessed June 18 2021].

NATIONAL CANCER INSTITUTE SEER PROGRAM. 2021. *Cancer Stat Facts: Female Breast Cancer* [Online]. Available: https://seer.cancer.gov/statfacts/html/breast.html [Accessed June 18 2021].

PATTERSON, R., PANTER, J., VAMOS, E. P., CUMMINS, S., MILLETT, C. & LAVERTY, A. A. 2020. Associations between commute mode and cardiovascular disease, cancer, and all-cause mortality, and cancer incidence, using linked Census data over 25 years in England and Wales: a cohort study. *The Lancet. Planetary health,* 4**,** e186-e194.

REMINGTON, P. L., CATLIN, B. B. & GENNUSO, K. P. 2015. The County Health Rankings: rationale and methods. *Population Health Metrics,* 13**,** 11.

ROQUETTE, R., PAINHO, M. & NUNES, B. 2017. Spatial epidemiology of cancer: a review of data sources, methods and risk factors. *Geospatial Health,* 12**,** 23-35.

SCHOOTMAN, M., GOMEZ, S. L., HENRY, K. A., PASKETT, E. D., ELLISON, G. L., OH, A., TAPLIN, S. H., TATALOVICH, Z. & BERRIGAN, D. A. 2017. Geospatial Approaches to Cancer Control and Population Sciences. *Cancer Epidemiology Biomarkers & Prevention,* 26**,** 472-475.

SCHOOTMAN, M., PEREZ, M., SCHOOTMAN, J. C., FU, Q., MCVAY, A., MARGENTHALER, J., COLDITZ, G. A., KREUTER, M. W. & JEFFE, D. B. 2020. Influence of built environment on quality of life changes in African-American patients with non-metastatic breast cancer. *Health & Place,* 63.

SCOTT, L., MOBLEY, L. R. & IL'YASOVA, D. 2017. Geospatial Analysis of Inflammatory Breast Cancer and Associated Community Characteristics in the United States. *International Journal of Environmental Research and Public Health,* 14.

SEILER, A., MURDOCK, K. W., GARCINI, L. M., CHIRINOS, D. A., RAMIREZ, J., JACKSON, B. & FAGUNDES, C. P. 2017. Racial/Ethnic Disparities in Breast Cancer Incidence, Risk Factors, Health Care Utilization, and Outcomes in the USA. *Current Breast Cancer Reports,* 9**,** 91-99.

TATALOVICH, Z., ZHU, L., ROLIN, A., LEWIS, D. R., HARLAN, L. C. & WINN, D. M. 2015. Geographic disparities in late stage breast cancer incidence: results from eight states in the United States. *International Journal of Health Geographics,* 14.

UNITED STATES CANCER STATISTICS 2018. USCS ASCII 1999-2015: ByCounty.

UNIVERSITY OF WISCONSIN POPULATION HEALTH INSTITUTE 2015. County Health Rankings & Roadmaps: 2015 Data Dictionary (PDF).

US CENSUS BUREAU 2012. 2010 Census.

US CENSUS BUREAU 2016. 2011-2015 American Community Survey 5-Year Estimates.

VIEIRA, V., WEBSTER, T., WEINBERG, J., ASCHENGRAU, A. & OZONOFF, D. 2002. Spatial analysis of lung, breast and colorectal cancer on Cape Cod using generalized additive modelling. *Epidemiology,* 13**,** S202-S202.

WANG, F. H., GUO, D. S. & MCLAFFERTY, S. 2012. Constructing geographic areas for cancer data analysis: A case study on late-stage breast cancer risk in Illinois. *Applied Geography,* 35**,** 1-11.

WEIR, H. K., THOMPSON, T. D., SOMAN, A., MOLLER, B. & LEADBETTER, S. 2015. The past, present, and future of cancer incidence in the United States: 1975 through 2020. *Cancer,* 121**,** 1827-1837.

ZAHND, W. E. & MCLAFFERTY, S. L. 2017. Contextual effects and cancer outcomes in the United States: a systematic review of characteristics in multilevel analyses. *Annals of Epidemiology,* 27**,** 739-748.

ZHAO, T., CUI, Z., MCCLELLAN, M. G., YU, D., AMY SANG, Q.-X. & ZHANG, J. 2020. Identifying county-level factors for female breast cancer incidence rate through a large-scale population study. *Applied Geography,* 125**,** 102324.

ZHU, H., LEI, X., FENG, J. & WANG, Y. 2012. Oral contraceptive use and risk of breast cancer: A meta-analysis of prospective cohort studies. *European Journal of Contraception and Reproductive Health Care,* 17**,** 402-414.

Appendix A. Variables and model coefficients in LASSO

| Non-modifiable | Modifiable | Expert | Features | Definition | Coefficient - Model 1 | Coefficient - Model 2 |
|---|---|---|---|---|---|---|
| | | | (Intercept) | (Intercept) | 119.534948 | 119.534948 |
| | | | Socioeconomics | | | |
| √ | | √ | Cen_API_perc | Percentage of Asian and Pacific Islander race population to the total population | 0.33229549 | |
| √ | | √ | Cen_Black_perc | Percentage of Black race population to the total population | 2.61080347 | 1.99433628 |
| √ | | √ | Cen_White_perc | Percentage of White race population to the total population | 3.12373969 | 1.78828625 |
| √ | | √ | Cen_Hispanic_perc | Percentage of Hispanic Ethnicity population to the total population | -2.64812125 | -2.70508618 |
| √ | | | Cen_Female_percent | Percentage of female gender among total population | | |
| √ | | | Cen_female_50nabove_perc | Percentage of female with age 50+ among all the female | | |
| √ | | | Cen_female_65nabove_perc | Percentage of female with age 65+ among all the female | | -0.92893743 |
| √ | | | Cen_50nabove_perc | Percentage of people with age of 50 and above | | |
| √ | | | Cen_65nabove_perc | Percentage of people with age of 65 and above | -0.69565771 | |
| √ | | | Cen_MedianAge_Female | Median age -- - Female | 1.68955630 | |
| √ | | | popUrban_percent | Percentage of urban population among total population (Urban:) | 1.90073454 | 1.56888438 |
| √ | | | popUrbanized_percent | Percentage of urbanized population among total population (Urban: - Inside urbanized areas) | 0.93209658 | 0.94161509 |
| √ | | √ | popRural_percent | Percentage of rural population among total population (Rural) | -0.13157845 | -0.05907354 |
| √ | | | pcinc | Estimate; Per capita income in the past 12 months (in 2015 Inflation-adjusted dollars) | | |
| √ | | √ | pov_percent | Percentage of families whose income below poverty level | -1.99611551 | -2.00811226 |
| √ | | | pov_hw_percpov | Percentage of poverty husband-wife families among all the poverty families | | |
| √ | | | pov_hw_percTotfam | Percentage of poverty husband-wife families among all the families | -0.75240383 | -0.19034772 |
| √ | | | pov_f_perc | Percentage of poverty female families among all the families | | |

| | | | | | |
|---|---|---|---|---|---|
| √ | | pov_f_percpov | Percentage of poverty female families among all the poverty families | 0.21559295 | 0.39528775 |
| √ | | pov_fc_perc | Percentage of poverty female families with children under 18 among all the families | | |
| √ | | pov_fcy_perc | Percentage of poverty female families with children under 5 among all the families | | |
| √ | | pov_fcyo_perc | Percentage of poverty female families with children under 5 & 5-17 among all the families | | |
| √ | | pov_fco_perc | Percentage of poverty female families with children 5-17 among all the families | | |
| √ | | pov_fnc_perc | Percentage of poverty female families with no child under 18 among all the families | -0.17874467 | -0.24780941 |
| √ | | Acs_edTotal_perc | Percentage of educated population among total population | | |
| √ | | Acs_edFemale_perc | Percentage of educated female among total population | | |
| √ | | Acs_edFemale_percF | Percentage of educated female among total female population | | |
| √ | | Acs_edFemale_percED | Percentage of educated female among total education population | 1.41734889 | 1.08550162 |
| √ | | flessHS_percF | Percentage of less than high school diploma female among total female population | -0.86102456 | |
| √ | | fHStoCA_percEDF | Percentage of female with high school to some college or associate's degree among total educated female population | | |
| √ | √ | fBachH_percEDF | Percentage of female with bachelor or higher degree among total educated female population | 0.87199472 | 0.76102073 |
| √ | √ | fMBSA_percF | Percentage of female with Management, business, science, and arts occupations | -0.18455638 | |
| √ | | fServ_percF | Percentage of female with Service occupations | | |
| √ | | fSO_percF | Percentage of female with Sales and office occupations | | |
| √ | | fNRCM_percF | Percentage of female with Natural resources, construction, and maintenance occupations | 0.90890420 | |
| √ | | fPTMM_percF | Percentage of female with Production, transportation, and material moving occupations | 0.24315493 | |
| √ | | medHHinc | Estimate; Median household income in the past 12 months (in 2015 Inflation-adjusted dollars) | | |
| √ | √ | medFinc | Estimate; Median family income in the past 12 months (in 2015 Inflation-adjusted dollars) | 0.70193039 | 0.33981419 |
| √ | | medNFHinc | Estimate; Median nonfamily household income in the past 12 months (in 2015 Inflation-adjusted dollars) | | |
| | | | Family Characteristics | | |
| √ | | FH_HW_percFH | Percentage of husband-wife family among all the FH | | |
| √ | √ | FH_HW_percTotHH | Percentage of husband-wife family among all the Households | | |
| √ | | FH_HW_c_percTotHH | Percentage of husband-wife family with own children among all the Households | | |
| √ | | FH_HW_cy_percTotHH | Percentage of husband-wife family with own children-under 6 years old among all the Households | 0.95115796 | |
| √ | | FH_HW_cyo_percTotHH | Percentage of husband-wife family with own children-under 6 and 6 to 17 years old among all the Households | 0.95115796 | -1.10746657 |
| √ | | FH_HW_co_percTotHH | Percentage of husband-wife family with own children-6 to 17 years old among all the Households | 2.47337808 | 2.27188708 |
| √ | | FH_HW_noc_percTotHH | Percentage of husband-wife family with no own children among all the Households | -1.33607665 | -1.40285982 |
| √ | | FH_F_percFH | Percentage of female householder among all the FH | | |
| √ | √ | FH_F_percTotHH | Percentage of female householder among all the Households | | |
| √ | | FH_F_c_percTotHH | Percentage of female householder with own children among all the Households | | |
| √ | | FH_F_cy_percTotHH | Percentage of female householder with own children-under 6 years old among all the Households | | |
| √ | | FH_F_cyo_percTotHH | Percentage of female householder with own children-under 6 and 6 to 17 years old among all the Households | | |
| √ | | FH_F_co_percTotHH | Percentage of female householder with own children-6 to 17 years old among all the Households | 3.53607197 | 2.53722369 |
| √ | | FH_F_noc_percTotHH | Percentage of female householder with no own children among all the Households | | |
| √ | √ | Child_bio_perc | Percentage of biological children that makes up family households among total population | -1.29335086 | |
| √ | | Child_adopted_perc | Percentage of adopted children that makes up family households among total population | -0.07363075 | -0.28771884 |
| √ | | Child_step_perc | Percentage of step children that makes up family households among total population | -3.24805095 | -2.81132053 |
| √ | | avgFS | Average family size -- - Total: | -0.34462761 | |
| √ | | avgFS18 | Average family size -- - Total: - 18 years and over | | |
| | √ | | avgHS | Average household size -- - Total: | |
| | √ | | avgHS18 | Average household size -- - Total: - 18 years and over | |
| | √ | | avgHSowner | Average household size -- - Owner occupied | |
| | √ | | avgHSrenter | Average household size -- - Renter occupied | |
| | colspan="5" Accessibility to Healthcare | | | | |
| | | √ | withIns_perc | Percentage of people with health insurance among all the population | |
| | | √ | Ins18_perc | Percentage of people of under 18 years with health insurance coverage among all under 18 years population | 0.40676330 |
| | √ | √ | Ins1864_perc | Percentage of people of 18 - 64 years with health insurance coverage among all 18 - 64 years population | -1.75100091 |
| | √ | √ | Ins65_perc | Percentage of people of 65 and over years with health insurance coverage among all 65 and over years population | 1.23571097 |
| | √ | √ | PCP.Rate | Rate of population to primary care physicians | 0.59228714 |
| | | √ | MHP.Rate | Rate of population to mental health providers | |
| | | √ | Medicare._Enrollees_perc | Percentage of No. of female Medicare Enrollees in total female population | 0.97029159 |
| | √ | √ | perc_Mammography | Percentage of female Medicare enrollees ages 67-69 that receive mammography screening | 1.45801124 |
| | | √ | pInsFemale | Percentage of Insured Female population | |
| | | √ | pInsedu25 | Percent Insured; Estimate; EDUCATIONAL ATTAINMENT - Civilian noninstitutionalized population 25 years and over | |
| | | √ | pInsLessHS | Percent Insured; Estimate; EDUCATIONAL ATTAINMENT - Civilian noninstitutionalized population 25 years and over - Less than high school graduate | |
| | | √ | pInsHS | Percent Insured; Estimate; EDUCATIONAL ATTAINMENT - Civilian noninstitutionalized population 25 years and over - High school graduate (includes equivalency) | 1.05319183 |
| | | √ | pInsSca | Percent Insured; Estimate; EDUCATIONAL ATTAINMENT - Civilian noninstitutionalized population 25 years and over - Some college or associate's degree | 1.64839440 |
| | | √ | pInsBach | Percent Insured; Estimate; EDUCATIONAL ATTAINMENT - | |
| √ | | pInsPOV138 | Civilian noninstitutionalized population 25 years and over - Bachelor's degree or higher Percent Insured; Estimate; RATIO OF INCOME TO POVERTY LEVEL IN THE PAST 12 MONTHS - Civilian noninstitutionalized population for whom poverty status is determined - Below 138 percent of the poverty threshold | | |
| √ | √ | perc_Diabetic | Percentage of diabetic Medicare enrollees receiving HbA1c test | | -0.06738325 |
| | | | Lifestyle | | |
| | √ | CHR_Fair.Poor_% | Percentage of adults reporting fair or poor health (age-adjusted) | | |
| | √ | Physically.Unhealthy.Days | Average number of physically unhealthy days reported in past 30 days (age-adjusted) | | |
| | √ | Mentally.Unhealthy.Days | Average number of mentally unhealthy days reported in past 30 days (age-adjusted) | | -0.65417455 |
| √ | √ | CHR_AdtSmoke_% | Percentage of adults who are current smokers | | 1.31499188 |
| √ | √ | CHR_AdtObes_% | Percentage of adults that report a BMI of 30 or more | | |
| | √ | Food.Environment.Index | Index of factors that contribute to a healthy food environment, 0 (worst) to 10 (best) | | |
| √ | √ | CHR_Physically.Inactive_% | Percentage of adults aged 20 and over reporting no leisure-time physical activity | | -1.58647504 |
| √ | √ | CHR_With.Access_% | Percentage of population with adequate access to locations for physical activity | | 1.24549849 |
| √ | √ | CHR_Excesdrink_% | Percentage of adults reporting binge or heavy drinking | | -1.41120058 |
| √ | | Association.Rate | Number of membership associations per 10,000 population | 0.86982334 | 1.02617952 |
| √ | | perc_Long.Commute...Drives.Alone | Among workers who commute in their car alone, the percentage that commute more than 30 minutes | 0.72211151 | 1.11420569 |
| | | | Environment | | |
| √ | √ | Average.Daily.PM2.5 | Average daily density of fine particulate matter in micrograms per cubic meter (PM2.5) | | -0.30251724 |
| √ | | perc_Pop.in.Viol | Percentage of population affected by a water violation among Total population with public water | | |
| √ | | perc_Food.Insecure | Percentage of food Insecure individuals among total population | | |
| √ | | perc_Limite | Percentage of population with Limited | | -0.33531016 |

|  |  | d.Access | access to healthy foods among total population |  |  |
| --- | --- | --- | --- | --- | --- |
| Total counts ||||||
| 64 | 29 | 23 |  |  | 31 | 38 |